%% file: ASPDAC27-FADiff.tex
\documentclass[sigconf,screen=true,anonymous=false,bookmarks=false,9pt]{acmart}
\copyrightyear{2027}
\acmYear{2027}
\setcopyright{cc}
\setcctype{by}
\acmConference[ASPDAC '27]{32nd Asia and South Pacific Design Automation Conference}{January 25--28, 2027}{Tokyo, Japan}
\acmBooktitle{32nd Asia and South Pacific Design Automation Conference (ASPDAC '27), January 25--28, 2027, Tokyo, Japan}
\acmDOI{10.1145/3840404.3848216}
\acmISBN{979-8-4007-2923-2/2027/01}

\input{setting-acm}

\graphicspath{{./figs/}}

\begin{document}

\title{
    FADiff: Fusion-Aware Differentiable Optimization for DNN Scheduling on Tensor Accelerators
}

\author{Shuao Jia}
\orcid{0009-0005-8753-4068}
\affiliation{%
    \institution{Beijing University of Posts and Telecommunications}
    \city{Beijing}
    \country{China}
}
\email{pdxqjiashuao@bupt.edu.cn}

\author{Zichao Ling}
\orcid{0009-0002-0942-6991}
\affiliation{%
    \institution{Beijing University of Posts and Telecommunications}
    \city{Beijing}
    \country{China}
}
\email{lingzichao@bupt.edu.cn}

\author{Chen Bai}
\orcid{0000-0002-1742-0090}
\affiliation{%
    \institution{Fudan University}
    \city{Shanghai}
    \country{China}
}
\email{cbai@fudan.edu.cn}

\author{Bei Yu}
\orcid{0000-0001-6406-4810}
\affiliation{%
    \institution{The Chinese University of Hong Kong}
    \city{Hong Kong}
    \country{China}
}
\email{byu@cse.cuhk.edu.hk}

\author{Kang Zhao}
\orcid{0000-0003-0502-8523}
\affiliation{%
    \institution{Beijing University of Posts and Telecommunications}
    \city{Beijing}
    \country{China}
}
\email{zhaokang@bupt.edu.cn}

\author{Jianwang Zhai}
\authornote{Corresponding author. This work is supported by the National Natural Science Foundation of China (No.~62404021), the Beijing Natural Science Foundation (No.~L2603021), and the Fundamental Research Funds for the BUPT (No.~2025AI4S20).}
\orcid{0000-0002-1581-3536}
\affiliation{%
    \institution{Beijing University of Posts and Telecommunications}
    \city{Beijing}
    \country{China}
}
\email{zhaijw@bupt.edu.cn}



\input{doc/abstract}

\maketitle
\pagestyle{plain}

\input{doc/intro}

\input{doc/prelim}

\input{doc/methodology}

\input{doc/result}

\input{doc/conclu}

\balance
\bibliographystyle{IEEEtran}
\bibliography{ref/Top, ref/intro}

\end{document}

%% file: setting-acm.tex
\iftrue
\fi

\usepackage{enumitem}
\setlist{leftmargin=5.08mm}

\iftrue
\fi

\usepackage{lipsum}
\usepackage{graphicx}
\usepackage{amsmath}
\usepackage{footnote}
\usepackage{mathtools}
\usepackage{mathrsfs}     
\usepackage{comment}
\usepackage{soul}                                          
\usepackage[subrefformat=parens,labelformat=parens]{subfig}
\usepackage{bm,bbm}
\usepackage{multirow}
\usepackage{threeparttable,booktabs}
\usepackage{blkarray}
\usepackage{tikz}
\usetikzlibrary{positioning,calc,fit,decorations.pathmorphing,shapes.geometric,shapes.gates.logic.US,calc}
\usetikzlibrary{arrows,arrows.meta,decorations.markings,shapes,shapes.arrows}
\usetikzlibrary{patterns,snakes}
\usetikzlibrary{decorations,decorations.pathreplacing}
\usetikzlibrary{backgrounds}
\usepackage{tikz-3dplot}
\usepackage{balance}
\usepackage{courier}                                       
\usepackage{cleveref}                                      
\usepackage[mathcal]{eucal}
\usepackage[noend]{algpseudocode}
\algrenewcommand\textproc{\texttt}
\makeatletter\let\float@addtolists\relax\makeatother
\usepackage{algorithm}
\usepackage{filecontents}                                  
\usepackage{pgfplots}
\usepackage{pgfplotstable}
\usepackage{pgf-pie}
\usepgfplotslibrary{groupplots}
\pgfplotsset{compat=newest}
\usepackage[figuresright]{rotating}
\usepackage{xcolor,colortbl}                               
\usepackage{adjustbox}                                     
\usepackage{makecell}
\usepackage{siunitx}
\usepackage{color}
\usepackage{diagbox}

\newcommand{\minisection}[1]{\vspace{.02in}\noindent{\textbf{#1}}.}

\theoremstyle{plain}

\theoremstyle{definition}

\algrenewcommand\textproc{\texttt}

\usepackage[skip=1pt]{caption}            
\definecolor{CUHKorange}{RGB}{244,106,18} 
\definecolor{CUHKblue}{RGB}{0,111,190}    
\definecolor{CUHKgreen}{RGB}{0,127,128}   
\definecolor{CUHKred}{RGB}{228,46,36}     
\definecolor{CUHKyellow}{RGB}{198,148,34} 
\definecolor{CUHKdark}{RGB}{114,44,114}   
\definecolor{CUHKmiddle}{RGB}{144,44,144} 
\definecolor{CUHKpurple}{RGB}{117,15,109} 
\definecolor{CUHKgold}{RGB}{221,163,0}    

\usepackage{tcolorbox}
\tcbuselibrary{skins,breakable}
    {\endtcolorbox}

%% file: doc/abstract.tex
\begin{abstract}

Efficiently deploying Deep Neural Networks (DNNs)—particularly Large Language Models (LLMs)—on tensor accelerators is essential for achieving high throughput and energy efficiency at the edge. 
However, the combinatorial explosion of the joint design space spanning intra-layer mapping and inter-layer fusion renders existing approaches inadequate. 
In this work, we present FADiff, a gradient-based optimization framework that automatically discovers optimal mapping and fusion strategies to accelerate DNN inference. 
We first formulate a unified, differentiable analytical cost model that accurately estimates the energy and latency of individual layer mappings as well as diverse fusion configurations. 
We then combine CP-SAT-based pre-enumeration of feasible discrete candidates with augmented-Lagrangian penalties over continuous relaxations, enabling a constrained gradient-based search that efficiently navigates the vast design space to identify optimal joint strategies. 
Experimental results show that FADiff consistently outperforms prior methods, delivering superior optimization quality with significantly faster convergence in both energy and latency.

\end{abstract}


%% file: doc/intro.tex
\section{Introduction}
\label{sec:introduction}


{E}{fficient} deployment of DNNs on resource-constrained hardware, such as edge tensor accelerators, is essential for realizing high performance on-device intelligence. 
A fundamental challenge in this endeavor is bridging the abstraction gap between high-level models and low-level hardware execution—a task that entails navigating a vast combinatorial design space in which two key decisions interact non-trivially: intra-layer mapping (how a single layer's computation and data are tiled and bound to hardware resources) and inter-layer fusion (how consecutive layers are grouped for joint execution). 
Together, these decisions govern data reuse, memory traffic, and ultimately energy and latency efficiency. 
As networks deepen and accelerator architectures grow more complex, this coupled mapping-fusion space becomes prohibitively difficult to explore and optimize.

Existing studies on DNN deployment design space exploration can be broadly categorized into heuristic and learning-based approaches. Heuristic frameworks such as TVM~\cite{chen2018tvm}, DeFiNES~\cite{mei2023defines}, SoMa~\cite{cai2025hpca}, and Klotski~\cite{bai2023klotski,bai2025klotskiv2} 
rely on combinatorial or search-based optimization to traverse the vast design space. 
Learning-based efforts, including Active Learning-guided optimization~\cite{sun2021active} and Deep Gaussian Transfer Learning~\cite{sun2022dgptl}, aim to improve efficiency through surrogate modeling and experience transfer. 
Despite their differences, both paradigms fundamentally depend on iterative surrogate evaluation and empirical search, resulting in slow convergence, limited interpretability, and a lack of theoretical grounding.

More recently, gradient-based optimization has emerged as an effective paradigm for accelerating DNN deployment design space exploration. 
The key insight, pioneered by DOSA~\cite{Hong2023micro} and Felix~\cite{zhao2024asplos}, lies in formulating differentiable energy and latency models. 
This transformation converts the discrete, combinatorial design space into a continuous domain, thereby enabling efficient gradient-based optimization. 
Instead of heuristic or learning-based algorithms, this paradigm uses analytical gradients in a theoretically sound way, enhancing optimization efficiency and interpretability by explicitly modeling parameter sensitivities.

\begin{figure}[t]
	\centering
	\includegraphics[width=0.8\linewidth]{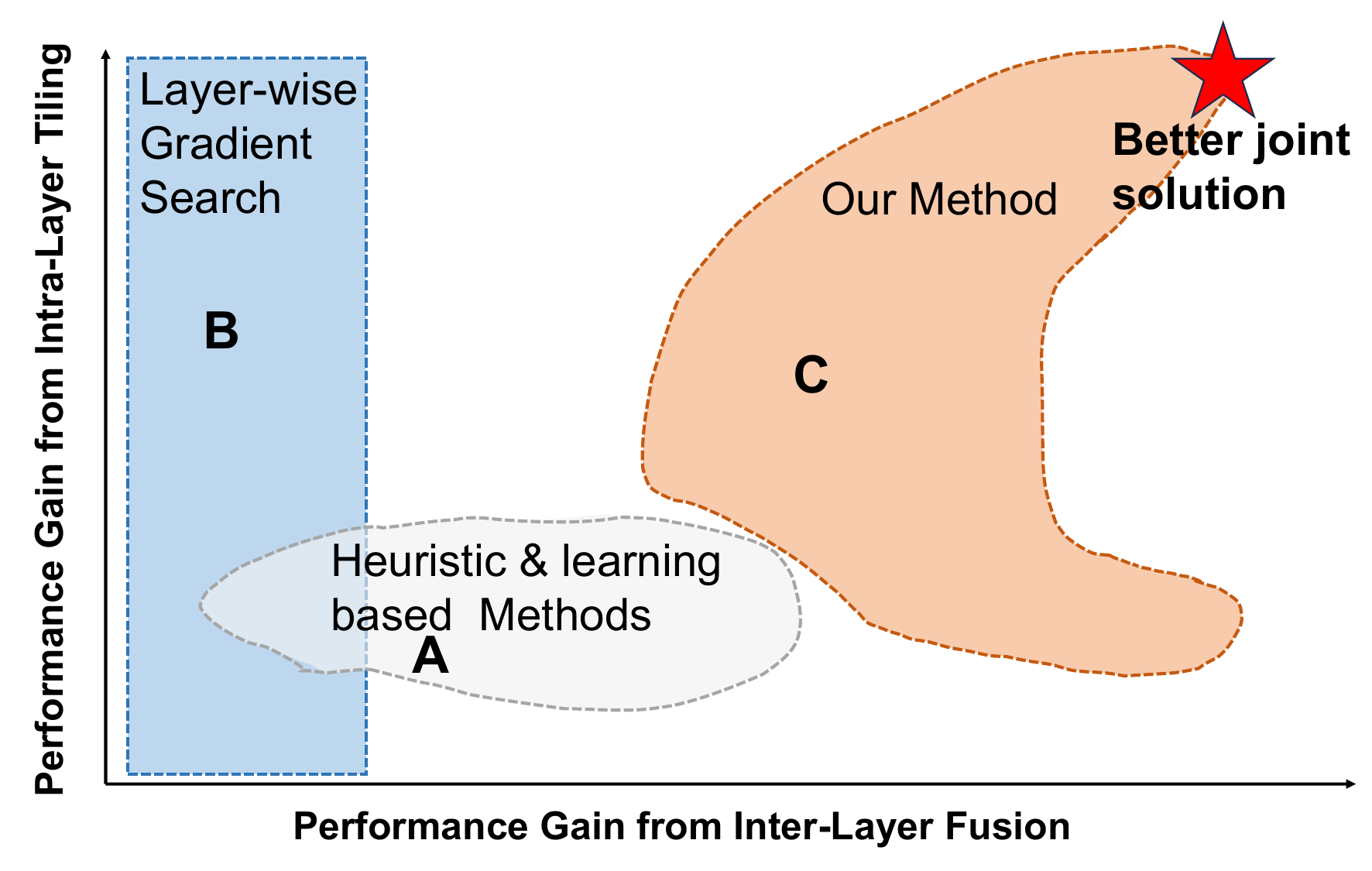} 
	\caption{
    The joint design space of inter-layer fusion and intra-layer mapping. 
    Traditional heuristic and learning-based methods (A) explore only a limited sub-region, while layer-wise gradient approaches (B) remain confined to a single axis. 
    In contrast, our method (C) performs a differentiable search over the full joint space, converging to a superior solution (red $\star$).
    }
	\label{fig1}
\end{figure}

However, these methods remain confined to layer-wise optimization and do not jointly address inter-layer fusion within the same differentiable framework.
Layer fusion---critical for reducing off-chip memory traffic---is still delegated to separate heuristic or learning-based passes, precluding truly global co-optimization of fusion structure and intra-layer mapping (\Cref{fig1}).
These shortcomings can be attributed to two fundamental causes.

Firstly, while general in principle, heuristic and learning-based methods face severe scalability barriers when addressing the combinatorially large, high-dimensional design space of joint mapping and fusion. 
For instance, Gaussian-process-based Bayesian Optimization (BO) requires inverting an $N \times N$ covariance matrix, incurring $O(N^3)$ cost in the number of evaluated samples $N$; in high-dimensional design spaces, $N$ must grow rapidly to explore the space, making BO quickly intractable.
Secondly, existing gradient-based approaches, though computationally efficient, are inherently limited to intra-layer optimization. 
Their fundamental constraint stems from the fact that inter-layer fusion decisions are inherently discrete—choosing which layers to merge admits no natural gradient. 
This discontinuity breaks the assumptions underlying standard backpropagation, forcing these methods to treat layers as independent and optimize each in isolation.

\begin{figure*}[!t]
    \centering
    \includegraphics[width=1\textwidth]{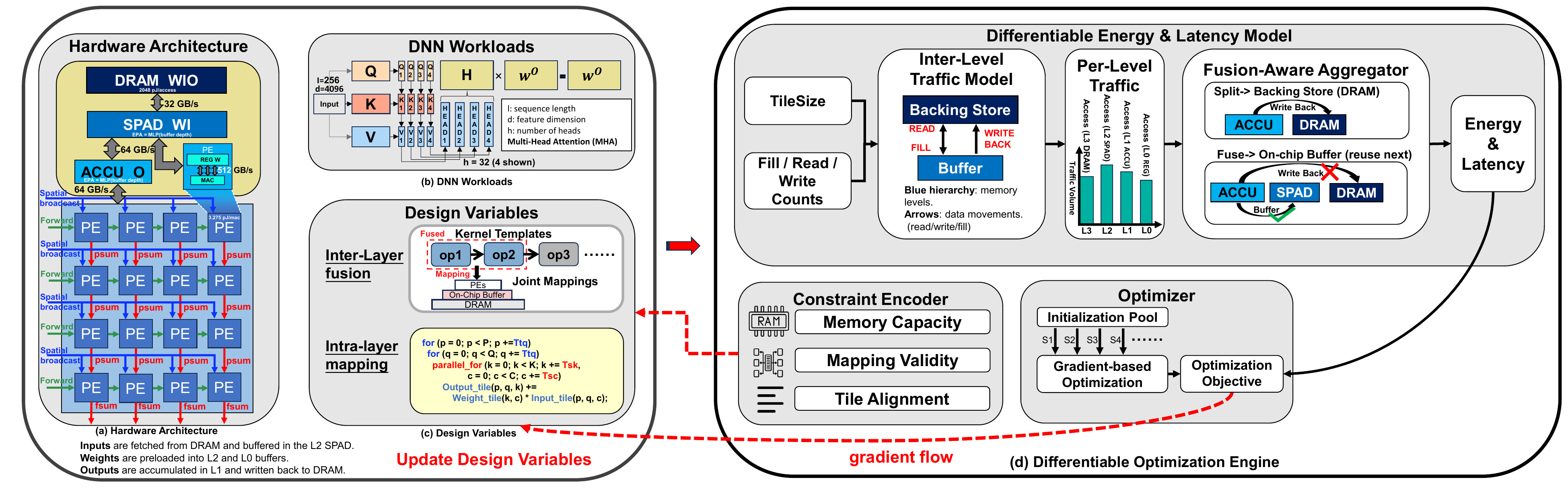}
    \caption{Overview of FADiff. 
    Left (regions~a--c): hardware architecture, DNN workloads, and design variables. 
    Right (region~d): differentiable optimization engine.}
    \label{fig2}
\end{figure*}


To bridge this gap, we propose FADiff, a gradient-based optimization framework that unifies intra-layer mapping and inter-layer fusion into a single, jointly differentiable optimization problem, as illustrated in \Cref{fig2}. 
At the heart of FADiff lies a differentiable analytical cost model that captures the complex interaction between mapping and fusion strategies in a unified form. 
This unified formulation enables efficient gradient-based co-optimization of both problems, yielding high-performance and hardware-valid DNN deployment solutions.
The main contributions are as follows:
\begin{itemize}
    \item We develop an analytical cost model that jointly captures intra-layer mapping and inter-layer fusion behaviors, enabling a unified differentiable formulation for energy and latency trade-offs on tensor accelerators.
    \item We embed validity constraints into the CP-SAT solver and add an augmented-Lagrangian penalty term to the objective, ensuring gradient-based optimization in continuous space with consistently feasible solutions.
    \item We demonstrate that FADiff achieves an average of $38.8\%$ reduction in Energy-Delay Product (EDP) compared with state-of-the-art optimization approaches.
\end{itemize}

The remainder of this paper is organized as follows.
\Cref{subsec:problem_formulation} formulates the joint mapping--fusion problem, \Cref{sec:method} details the proposed FADiff framework, \Cref{sec:evaluation} presents the experimental results, and \Cref{sec:conclusion} concludes the paper.

%% file: doc/prelim.tex
\section{Problem Formulation}
\label{subsec:problem_formulation}
In this section, we formalize the joint mapping--fusion deployment problem addressed in this work.

We model the DNN as a directed acyclic graph (DAG), i.e., $G = (V, E)$, where each vertex $v \in V$ represents a computational layer and each edge $e \in E$ encodes a tensor dependency, thereby capturing the producer-consumer relationships that govern fusion legality.
A \emph{deployment strategy} specifies the complete spatio-temporal execution of the network on the target hardware through intra-layer mapping and inter-layer fusion.
Our objective is to identify the strategy that minimizes the EDP subject to on-chip buffer capacity and dataflow constraints.


%% file: doc/methodology.tex
\section{Methodology}
\label{sec:method}

Existing deployment frameworks for tensor accelerators face a fundamental trade-off: non-differentiable combinatorial search incurs prohibitive cost, while differentiable alternatives remain confined to layer-wise mapping and neglect fusion entirely.
As illustrated in~\Cref{fig2}, we resolve this tension with a unified differentiable framework that co-optimizes layer mapping and fusion end-to-end.

In this section, we present the continuous representation of strategies in Section~\ref{sec:repr}, followed by the differentiable energy and latency model in Section~\ref{sec:model}, the differentiable modeling of optimization constraints in Section~\ref{sec:constraint}, and the end-to-end optimization framework in Section~\ref{sec:e2e}.

\subsection{Continuous Representation of Strategies}
\label{subsec:continuous_repr}
\label{sec:repr}

The central obstacle to gradient-based optimization is the discrete nature of mapping and fusion decisions.
We overcome this by introducing continuous parameterizations that relax both strategy spaces into a jointly differentiable formulation.

\minisection{Mapping Representation}
\label{subsec:mapping_repr}
Each layer's mapping is fully characterized by its tiling factors across all problem dimensions and memory levels.
We adopt a 7-dimensional problem space $\{N, K, C, P, Q, R, S\}$, where $N$ is batch size, $K$ and $C$ are output and input channel counts, $P$ and $Q$ are output spatial dimensions, and $R$ and $S$ are kernel height and width.
This formulation subsumes both CONV and GEMM (the latter setting $P{=}Q{=}R{=}S{=}1$). 
Tiling targets a 4-level memory hierarchy: $L3$ (DRAM), $L2$ (scratchpad), $L1$ (accumulator), and $L0$ (PE registers).
We denote the temporal tiling factor of dimension $d$ at level $m$ by $T_{t,d,m}$, and the corresponding spatial factor by $T_{s,d,m}$; where unambiguous, we write $T_{d,m}$ to refer to either.

Although gradient-based optimization requires continuous tiling factors $\{T_{t,d,m}\}$ and $\{T_{s,d,m}\}$, the design space is inherently discrete.
We therefore optimize continuous relaxations at all memory levels except the outermost, whose factors are derived from the total dimension size to guarantee that the product along each dimension does not exceed it.
An augmented-Lagrangian penalty ensures $T_{d,m} \geq 1$ (\Cref{subsec:constraints}).
To reconcile continuous optimization with discrete feasibility, we pre-enumerate all valid mapping candidates per layer using the CP-SAT solver and periodically project each relaxed factor onto its nearest valid candidate, maintaining consistency with the discrete space throughout training.

\begin{figure}[!t]
    \centering
    \includegraphics[width=0.95\columnwidth]{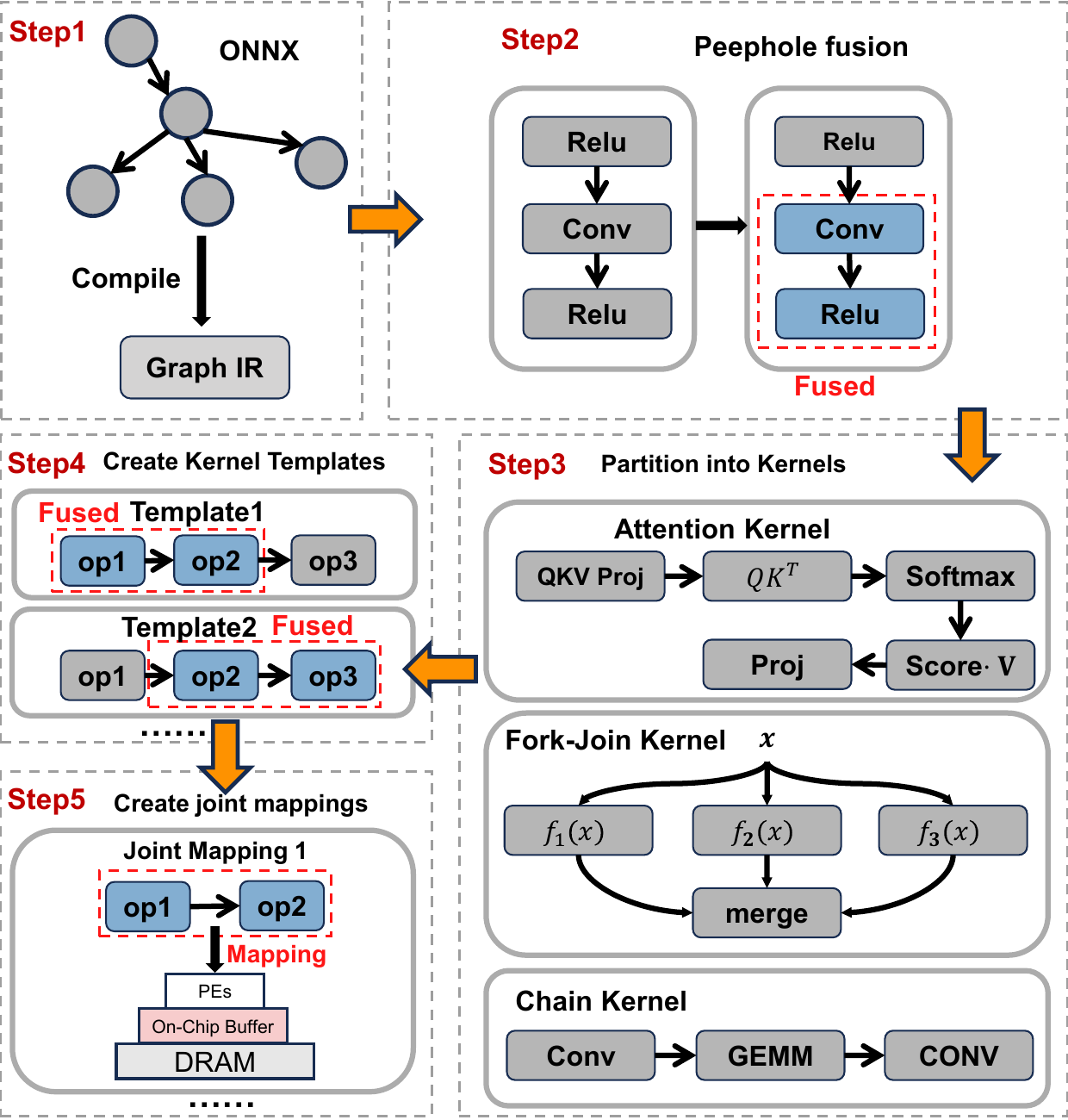}
    \caption{Overview of the five-step pipeline for generating the layer fusion design space.}
    \label{fig:fig6}
\end{figure}

\minisection{Kernel-Template-Based Layer Fusion Representation}
Existing gradient-based methods optimize intra-layer mapping effectively but do not model cross-layer fusion. 
We introduce a differentiable, template-based fusion representation that captures the topology of modern DNNs while remaining compatible with layer-wise mapping optimization.

As shown in Fig.~\ref{fig:fig6}, we compile the model's ONNX graph into a graph-level IR, apply peephole fusion of lightweight operators (e.g., ReLU, Clip) into preceding compute-bound layers, and partition the result into optimizable subgraphs called \textit{kernels}.
Representative kernel topologies include \textit{attention} (Q/K projection $\rightarrow$ score MatMul $\rightarrow$ Softmax $\rightarrow$ V MatMul), \textit{residual} (two branches merging at Add), \textit{fork-join} (a shared ancestor splitting into paths that reconverge at Concat or Mul), and \textit{chain} (a linear sequence of Conv/GEMM/MatMul).

For each kernel $k \in \mathcal{K}$, we construct a finite, size-tunable template library $\mathcal{T}_k$. 
Each template $\tau \in \mathcal{T}k$ encodes (1) fused operator groups $\mathcal{G}{k,\tau}$, whose depth determines the intermediate buffer footprint; (2) tile-alignment constraints between producers and consumers (\Cref{subsubsec:tile-alignment}); and (3) intermediate-tensor residency constraints that retain data on-chip (\Cref{subsubsec:memory-capacity}). 
All templates are generated by a CP-SAT solver, guaranteeing correctness and hardware-constraint satisfaction by construction.

Within each fused group $G \in \mathcal{G}_{k,\tau}$, the constituent nodes are jointly mapped: their combined tiling simultaneously determines buffered tile sizes, producer-consumer alignment, and on-chip residency—all of which must hold for the fusion to be valid.
We use the CP-SAT solver to pre-enumerate per-node mapping candidates under these hard constraints; the subsequent optimization over this candidate pool is described in~\Cref{end_2_end}.

Layer fusion thus reduces to a two-level problem: discrete template selection per kernel, followed by continuous joint-mapping optimization within each fused group.

Template selection is made differentiable via Gumbel-Softmax with straight-through estimation (STE). 
Learnable logits $l_{k,\tau}$ yield the relaxed categorical distribution:
\begin{equation}
p_{k,\tau}
=
\frac{\exp\!\big((l_{k,\tau}+g_{k,\tau})/\lambda\big)}
{\sum_{\tau' \in \mathcal{T}_k}\exp\!\big((l_{k,\tau'}+g_{k,\tau'})/\lambda\big)},
\end{equation}
where $g_{k,\tau}\sim \mathrm{Gumbel}(0,1)$ and $\lambda$ is the temperature.

For the joint mapping within each fused group, we optimize continuous factors—consistent with the standalone mapping formulation—while enforcing feasibility through augmented-Lagrangian penalties (\Cref{subsec:constraints}) and periodic projection onto the pre-enumerated candidate set (\Cref{end_2_end}).

\minisection{Unified Hierarchical Representation}
FADiff parameterizes the design space with two coupled groups.
\emph{Continuous mapping factors} $\{T_{t,d,m}\}$ and $\{T_{s,d,m}\}$ govern tiling and spatial-unrolling decisions for every layer across the memory hierarchy; the same parameterization applies to both standalone and fused layers, with feasibility maintained through augmented-Lagrangian penalties and periodic rounding.
\emph{Discrete fusion logits} $\{l_{k,\tau}\}$ select, for each detected kernel $k\in\mathcal{K}$, a fusion template specifying its constituent layers and pipeline depth.
Formally, the complete parameter set is:
\begin{equation}
\Theta = \bigl\{\,T_{t,d,m}^{(l)},\, T_{s,d,m}^{(l)}\bigr\}
\;\cup\;
\bigl\{\,l_{k,\tau} \mid k \in \mathcal{K},\; \tau \in \mathcal{T}_k\bigr\}.
\end{equation}

Both groups are jointly optimized: mapping factors via gradient descent on continuous relaxations (Section~\ref{subsec:mapping_repr}), template selection via Gumbel-STE over a CP-SAT--pre-enumerated feasible set---enabling FADiff to co-optimize mapping and fusion in a single differentiable pass.

\subsection{Differentiable Energy and Latency Model}
\label{sec:model}
\label{subsec:modeling}

We develop a unified differentiable cost model that estimates end-to-end energy and latency for a complete deployment strategy, enabling gradient-based exploration of the joint design space.

\minisection{Data Traffic}
Data movement dominates both latency and energy in tensor accelerators, often rivaling computation cost~\cite{Huang2024isca}.
We formulate total inter-level data traffic as a differentiable function of continuous mapping parameters and model layer fusion's impact on on-chip reuse via the kernel-template representation, ensuring gradient propagation through the full memory hierarchy.

\noindent
\textbf{(1) Fill Traffic.}\;
For tensor $T \in \{I, W\}$, the fill traffic into level $L_i$ is:
\begin{align}
\mathrm{Fill}(L_i, T)
= \mathrm{TileSize}(i,T) \cdot \mathrm{FetchCount}(i,T),
\end{align}
\begin{align}
\mathrm{TileSize}(i,T) &= 
\prod_{d \in \mathcal{D}(T)} 
\prod_{k=0}^{i} T_{t,d,k} \cdot T_{s,d,k},
\label{eq:tilesize} \\[4pt]
\mathrm{FetchCount}(i,T) &=
\prod_{d \in \mathcal{D}(T)} 
\prod_{k=i+1}^{M} 
T_{t,d,k},
\label{eq:fetchcount}
\end{align}
where $\mathcal{D}(T)$ denotes the set of problem dimensions associated with $T$ (e.g., $\{R,S,C,K\}$ for weights) and $L_M$ is the outermost level (DRAM).

\noindent
\textbf{(2) Read Traffic.}\;
Read traffic captures data transferred from a lower (slower) to a higher (faster) memory level upon request. 
We distinguish two types: \emph{inter-memory reads}, which move entire tiles between levels, and \emph{PE-supplying reads}, which feed data directly to the compute array.

\noindent
\textit{(a) Inter-Memory Reads.}\;
For tensor $T \in \{I, W\}$, the read traffic from $L_{i+1}$ is:
\begin{align}
    \mathrm{Read}(L_{i+1}, T)
    &=
    \mathrm{TileSize}(i,T) \cdot \mathrm{FetchCount}(i,T).
\end{align}
No broadcast reuse applies, as no broadcast network exists between memory levels.

\noindent
\textit{(b) PE-Supplying Reads.}\;
Reads from the innermost buffer $L_{\mathrm{i}}$ feed the PE array. 
The traffic equals total MACs divided by the spatial broadcast factor:
\begin{align}
    \mathrm{Read}(L_{\mathrm{i}}, T)
    &=
    \frac{\mathrm{Ops}}{\mathrm{Bcast}_T},
    \qquad
    \mathrm{Bcast}_T = \prod_{d \notin \mathcal{D}(T)} T_{s,d,m},
\end{align}
where $\mathrm{Ops}$ is the layer's MAC count and $\mathrm{Bcast}_T$ captures multicast reuse over dimensions not indexed by $T$.

\noindent
\textbf{(3) Write-Back Traffic.}\;
Output tensors and partial sums reside only in the L1 accumulator and L3 DRAM (bypassing L2/L0), yielding two write-back types.

\noindent
\textit{(a) Inter-Memory Write-back.}\;
Completed output tiles drain from L1 to DRAM:
\begin{align}
\mathrm{WriteBack}(L_i, T)
&=
\mathrm{TileSize}(i,T)\cdot\mathrm{WriteCount}(i,T),
\label{eq:writeback}
\end{align}
where $\mathrm{WriteCount}(i,T)$ is determined by the outer temporal loop iterations, analogous to $\mathrm{FetchCount}$ in~\Cref{eq:fetchcount}.

\noindent
\textit{(b) Accumulation Write-back.}\;
Partial sums are spatially reduced within the PE array before writing to L1:
\begin{align}
    \mathrm{WriteBack}(L_{\mathrm{1}}, T)
    &=
    \frac{\mathrm{Ops}}{\mathrm{Reduce}_T},
    \qquad
    \mathrm{Reduce}_T = \prod_{d \notin \mathcal{D}(T)} T_{s,d,m}.
\end{align}

\minisection{Latency Model}
We adopt a roofline-style model assuming full overlap of computation and memory access.
Each layer's latency is the maximum of compute and memory-bound terms; total latency sums over all layers:
\begin{align}
\label{eq:latency}
\mathrm{Latency}(l_{i'}) = 
\max\!\left(
\frac{\mathrm{Ops}}{\mathrm{PEs}},\;
\max_i \frac{\mathrm{Access}(L_i)}{BW_i}
\right),
\end{align}
where $\mathrm{PEs}$ is the effective PE count (determined by $T_{s,d,k}$), $\mathrm{Access}(L_i)$ aggregates fill, read, and write-back traffic at level $L_i$, and $BW_i$ is its effective bandwidth (\Cref{fig2}(a)).

\minisection{Energy Model}
The per-layer energy sums computation and data-movement contributions; total model energy aggregates across layers:
\begin{align}
E_{\text{compute}} 
&= \mathrm{Ops}\cdot \mathrm{EnergyPerOp}, \\[4pt]
E_{\text{datamove}}
&= \sum_{i} 
\mathrm{Access}(L_{i}) 
\cdot 
\mathrm{EPA}_i,
\end{align}
where $\mathrm{EnergyPerOp}$ and $\mathrm{EPA}_i$ are listed in~\Cref{fig2}(a).
The optimization objective is the EDP.
Because all model components are constructed differentiably, gradients of the EDP with respect to mapping parameters $T_{t,d,m}$, $T_{s,d,m}$ and fusion logits $\{l_{k,\tau}\}_{\tau \in \mathcal{T}_k}$ are obtained directly via automatic differentiation.

\subsection{Differentiable Modeling of Optimization Constraints}
\label{sec:constraint}
\label{subsec:constraints}

We enforce constraints through a two-pronged strategy.
\emph{Hard} constraints---mapping validity, memory capacity, and adjacent-tile alignment---are embedded in the CP-SAT solver during pre-enumeration, ensuring every discrete candidate is feasible by construction.
\emph{Soft} augmented-Lagrangian penalties provide differentiable guidance between projection steps, discouraging infeasible drift of the continuous trajectory.
Because mapping variables are periodically projected onto the nearest valid candidate and the final solution is always evaluated post-projection, feasibility is guaranteed regardless of intermediate relaxation.

\minisection{Mapping-Validity Constraints}
Mapping validity enforces legal temporal and spatial tiling via three conditions. Each factor must satisfy $T_{d,m} \geq 1$ to exclude degenerate mappings. 
Spatial parallelism is bounded by the array size:
\begin{align}
\prod_{d} T_{s,d,m} \leq N_{\mathrm{PE}},
\end{align}
where $m$ is the memory level attached to the PE array. 
Finally, temporal and spatial factors must exactly tile each dimension:
\begin{align}
\prod_{m} T_{t,d,m} \cdot T_{s,d,m} = D_d, \quad \forall\, d.
\end{align}

\minisection{Memory-Capacity Constraints}
\label{subsubsec:memory-capacity}
For each fused group $G \in \mathcal{G}_{k,\tau}$, the weights and activations of all constituent nodes must co-reside in the on-chip buffer. The aggregate storage at level $i$ is:
\begin{align}
    \mathrm{SizeReq}(G, i) = \sum_{v \in G} \bigl( S_{W}(v, i) + S_{I}(v, i) \bigr) \leq C_i, \quad \forall\, G,\, i,
\end{align}
where $S_{W}$ and $S_{I}$ are the per-node weight and activation footprints derived from~\Cref{eq:tilesize}, and $C_i$ is the physical capacity of level $i$.

\minisection{Adjacent-Tile Alignment Constraints}
\label{subsubsec:tile-alignment}
Within each fused group $G \in \mathcal{G}_{k,\tau}$, the output tile of layer $v_i$ must match the input tile of its successor to ensure correct inter-layer data reuse:
\begin{align}
    \mathbf{o}_{v_i} = \mathbf{i}_{v_{i+1}},
    \qquad
    \forall\,(v_i, v_{i+1}) \in G,\ \forall\, G \in \mathcal{G}_{k,\tau},
\end{align}
where $\mathbf{o}_{v_i}{=}(o_p, o_q, o_k)$ and $\mathbf{i}_{v_{i+1}}{=}(i_h, i_w, i_c)$ are the output and input tile shapes, respectively.

All three constraint classes are enforced dually: as augmented-Lagrangian penalty terms that provide differentiable guidance during continuous optimization, and as hard constraints in the CP-SAT solver that guarantee the pre-enumerated candidate set is exclusively feasible. 
Periodic projection of mapping variables onto their nearest valid candidate maintains feasibility throughout.

\begin{algorithm}[t]
\caption{Multi-Start End-to-End Optimization}
\label{alg:multi_start_e2e}
\begin{algorithmic}[1]
\Require $\Theta$, $\mathcal{K}$, $N$, $R$, $\Delta$
\Ensure $\Theta^\star$

\State $\{\Theta_n^{(0)}\}_{n=1}^{N} \leftarrow \textsc{MaximinInit}(N)$
\State $\Theta^\star \leftarrow \varnothing,\quad \mathcal{J}^\star \leftarrow +\infty$
\For{$n = 1$ to $N$}
    \State $\Theta \leftarrow \Theta_n^{(0)}$
    \For{$r = 1$ to $R$}
        \State sample $\{\tau_k\}_{k \in \mathcal{K}}$ from $\{l_{k,\tau}\}$ via STE
        \State update mapping factors by gradient descent
        \If{$r \bmod \Delta = 0$}
            \State project mapping factors to nearest valid candidates
        \EndIf
        \If{converged} \textbf{break} \EndIf
    \EndFor
    \State project mapping factors to valid candidates
    \If{$\mathcal{J}(\Theta) < \mathcal{J}^\star$}
        \State $\Theta^\star \leftarrow \Theta,\quad \mathcal{J}^\star \leftarrow \mathcal{J}(\Theta)$
    \EndIf
\EndFor
\State \Return $\Theta^\star$
\end{algorithmic}
\end{algorithm}

\subsection{End-to-End Optimization Framework}
\label{sec:e2e}
\label{end_2_end}
We solve the joint problem via multi-start gradient descent (\Cref{alg:multi_start_e2e}).
$N$ trajectories are initialized with a maximin criterion over the full parameter set $\Theta$ (\Cref{subsec:continuous_repr}) to improve coverage of the non-convex landscape.

Continuous mapping factors for both standalone and fused-group layers are optimized directly and projected every $\Delta$ iterations onto the nearest valid candidate in the pre-enumerated set---avoiding the scalability issues of direct discrete selection over large mapping spaces (\Cref{relaxation_ablation}).
Kernel-template selection is handled via Gumbel-STE: hard discrete choices in the forward pass, soft relaxation in the backward pass.
Throughout, augmented Lagrangian penalties (\Cref{subsec:constraints}) steer trajectories toward feasible, high-quality regions so that final projections yield strong---not merely legal---solutions.
After convergence, all factors are discretized and the best trajectory is returned as $\Theta^\star$.


%% file: doc/result.tex
\section{Evaluation}
\label{sec:evaluation}
In this section, we present the experimental setup in Section~\ref{subsec:setup}, followed by the validation of the differentiable cost models in Section~\ref{subsec:validation}, the comparison of optimization quality in Section~\ref{subsec:baseline}, and the ablation study in Section~\ref{subsec:ablation}.

\begin{table*}[t]
\centering
\caption{EDP Comparison Across Methods and Gemmini Configurations.}
\label{tab:raw_edp_all_methods}
\resizebox{\textwidth}{!}{%
\renewcommand{\arraystretch}{1.2}
\begin{tabular}{lcccccccccc}
\toprule
\multirow{2}{*}{\textbf{Model}} & \multicolumn{5}{c}{\textbf{Gemmini large}} & \multicolumn{5}{c}{\textbf{Gemmini small}} \\    
\cmidrule(lr){2-6} \cmidrule(lr){7-11}
& \textbf{RD} & \textbf{GA} & \textbf{BO} & \textbf{DOSA~\cite{Hong2023micro}} & \textbf{FADiff}
& \textbf{RD} & \textbf{GA} & \textbf{BO} & \textbf{DOSA~\cite{Hong2023micro}} & \textbf{FADiff} \\
\midrule
VGG19       & 3.890e+13 & 2.100e+13 & 4.200e+13 & 4.540e+13 & \textbf{1.730e+13} & 1.510e+14 & 9.060e+13 & 2.030e+14 & 7.380e+13 & \textbf{4.710e+12} \\
VGG16       & 2.560e+13 & 1.780e+13 & 2.830e+13 & 1.610e+13 & \textbf{1.210e+13} & 1.040e+14 & 4.930e+13 & 7.860e+13 & 1.700e+13 & \textbf{3.240e+12} \\
MobileNetV1 & 3.760e+12 & 1.670e+12 & 2.010e+12 & 3.230e+12 & \textbf{1.600e+12} & 7.200e+12 & 4.810e+12 & 6.690e+12 & 1.030e+13 & \textbf{3.620e+12} \\
GPT3-6.7B   & 2.170e+14 & 7.920e+13 & 1.110e+14 & 8.330e+13 & \textbf{4.810e+13} & 1.530e+15 & 1.170e+15 & 1.140e+15 & 1.910e+15 & \textbf{3.950e+14} \\
ResNet18    & 6.570e+11 & 4.300e+11 & 1.130e+12 & 5.030e+11 & \textbf{2.740e+11} & 1.080e+12 & 5.710e+11 & 1.500e+12 & 8.600e+11 & \textbf{5.640e+11} \\
\midrule
\textbf{Average} & 5.718e+13 & 2.402e+13 & 3.689e+13 & 2.971e+13 & \textbf{1.587e+13} & 3.587e+14 & 2.631e+14 & 2.860e+14 & 4.024e+14 & \textbf{8.143e+13} \\
\bottomrule
\end{tabular}%
}
\end{table*}

\subsection{Experimental Setup}
\label{subsec:setup}

All experiments run on a dual-socket Intel\textsuperscript{\textregistered} Xeon\textsuperscript{\texttrademark} Platinum~8375C server with 512\,GB DRAM.
We target Gemmini~\cite{Genc2021dac} under two configurations that span a representative embedded design space: a \emph{large} variant ($32\times32$ systolic array, 64\,KB L1, 512\,KB L2) and a \emph{small} variant ($16\times16$, 8\,KB L1/L2), enabling us to characterize scaling behavior across distinct compute-to-memory ratios.
Memory bandwidth, DRAM energy-per-access (EPA), and MAC energy are adopted from DOSA~\cite{Hong2023micro}; on-chip buffer EPA is derived from an MLP surrogate fitted to CACTI~\cite{shafaei2014fincacti} data.
All energy and area parameters are summarized in~\Cref{fig2}(a).
We report energy-delay product (EDP) as the primary deployment-quality metric.


\subsection{Validation of Differentiable Cost Models}
\label{subsec:validation}

We first validate single-layer accuracy against Timeloop~\cite{Parashar2019ispass} / Accelergy~\cite{wu2019iccad} across standard, depthwise, pointwise, large-kernel convolutions, and fully connected layers. 
The model achieves 96\% accuracy in predicting memory access counts, perfect rank correlation for latency (Kendall's $\tau{=}1.0$, Spearman's $\rho{=}1.0$), and strong rank correlation for energy ($\tau{=}0.78$, $\rho{=}0.92$).

\begin{figure}[t]
  \centering
  \includegraphics[width=\linewidth]{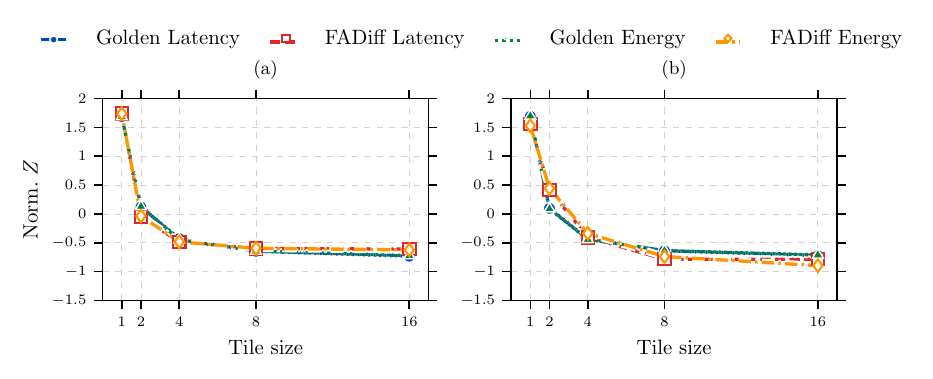}
  \caption{%
    Comparison of Z-score normalized trends under different tile sizes.
    Left: two-layer fusion. Right: three-layer fusion.
  }
  \label{fig:fig3}
\end{figure}

We then assess multi-layer fused operators against DeFiNES~\cite{mei2023defines}, whose analytical model is calibrated to silicon. 
Because absolute values are sensitive to process and microarchitectural assumptions, we compare only ranking consistency. 
As shown in~\Cref{fig:fig3}, Z-score--normalized latency and energy trends closely track the DeFiNES reference for both two- and three-layer fusion, confirming that our differentiable cost model preserves the relative ordering required for reliable gradient-based optimization.


\subsection{Comparison of Optimization Quality}
\label{subsec:baseline}

\minisection{Methodological Baselines}
To isolate the benefit of gradient-based search in this high-dimensional space, we compare against a Genetic Algorithm (GA)~\cite{holland1975adaptation} and Bayesian Optimization (BO)~\cite{snoek2012practical}, both operating on the identical differentiable cost model and candidate space.
All experiments use the Large-Gemmini configuration (\Cref{subsec:setup}).
FADiff launches $N{=}20$ independent trajectories of $R{=}1{,}000$ gradient steps with projection interval $\Delta{=}100$; the Gumbel temperature anneals from $\lambda{=}8{\to}1$ (phase~1) and $1{\to}0.1$ (phase~2).
Candidate pools are capped at 10{,}000 (standalone) and 30{,}000 (fused group) entries to balance enumeration cost against feasible-space coverage.
All methods share the same total evaluation budget unless otherwise noted.

\begin{figure}[t]
  \centering
  \includegraphics[width=1.0\linewidth]{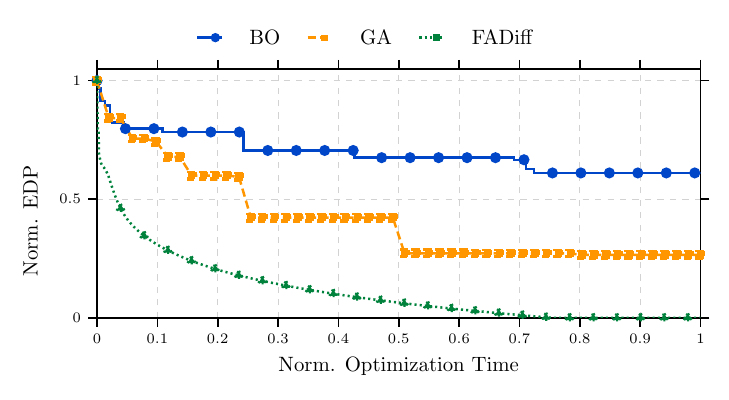}
  \caption{
  EDP vs. optimization time for GA, BO, and our gradient-based method; with the same optimization budgets, our method converges faster to lower EDP.
  }
  \label{fig:fig4}
\end{figure}

As shown in~\Cref{fig:fig4}, our gradient-based method achieves substantially lower EDP in far less wall-clock time than GA and BO under an equal budget---underscoring the efficiency of analytical gradients over heuristic or surrogate-based search in high-dimensional mapping spaces.
To contextualize these dynamics within end-to-end deployment quality, we additionally run Random Search (RD), GA, and BO on the full model suite; their best EDP is reported alongside DOSA~\cite{Hong2023micro} and our method in~\Cref{tab:raw_edp_all_methods}.

\minisection{Differentiable Framework Baselines}
\label{subsubsec:differentiable_framework_baseline}
We compare FADiff against DOSA~\cite{Hong2023micro}, which shares our gradient-based paradigm but optimizes each layer independently without modeling fusion.
Both methods are evaluated on the two Gemmini configurations (\Cref{subsec:setup}) under equal budget.
The workload suite spans representative CNNs~\cite{he2016cvpr, howard2017mobilenetsefficientconvolutionalneural, simonyan2015deepconvolutionalnetworkslargescale} and a GPT-3-6.7B decoder-only Transformer~\cite{mann2020language}, for which we optimize MHA~\cite{vaswani2017attention} and FFN blocks (tensor dimensions in~\Cref{fig2}(b); FFN hidden size 16{,}384).

As shown in~\Cref{tab:raw_edp_all_methods}, FADiff achieves the lowest EDP across all ten workload--hardware combinations, reducing EDP by \textbf{38.8\%} on average relative to the best competing baseline.
Gains are especially pronounced on Small-Gemmini, indicating that joint fusion-and-mapping optimization becomes increasingly critical under tight resource budgets.
The underlying design space is astronomically large---rendering RD intractable, BO prohibitively expensive, and GA prone to premature convergence.
By unifying fusion structure and per-layer mapping in a single differentiable formulation, FADiff enables gradient-driven search to efficiently navigate this otherwise intractable space.

\subsection{Ablation Study}
\label{subsec:ablation}

\begin{figure}[!t]
    \centering
    \includegraphics[width=1.0\columnwidth]{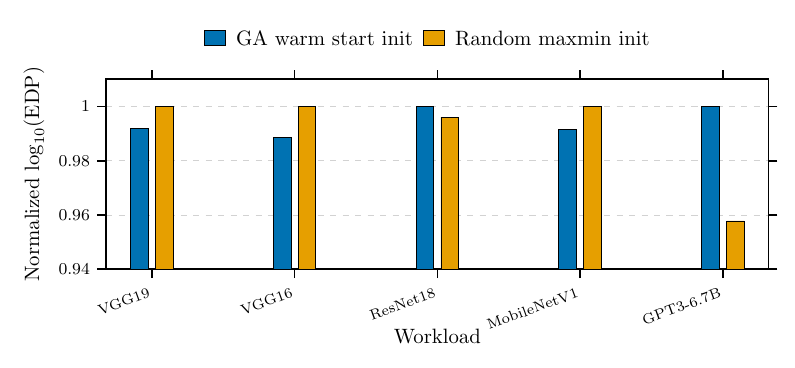}
    \caption{Effect of initialization strategy on optimization quality. GA warm-start and random (maximin) initialization are compared in terms of final EDP.}
    \label{fig:fig9}
\end{figure}

\minisection{Effect of Initialization Strategy} 
We compare two strategies under an equal evaluation budget: (1) random initialization from 100 maximin-sampled points, and (2) GA warm-start from a single elite individual after 10 generations (population 10). 
\Cref{fig:fig9} reports EDP on the Large Gemmini across all five workloads. 
GA warm-start consistently outperforms on structurally regular CNNs (MobileNetV1, VGG16, VGG19), where brief evolutionary pre-exploration steers optimization toward favorable basins. 
Conversely, random initialization prevails on GPT-3, whose large, strongly coupled design space is sensitive to the premature bias introduced by GA seeding. 
ResNet18 exhibits intermediate behavior, with neither strategy dominating.

\begin{figure}[!t]
    \centering
    \includegraphics[width=1.0\columnwidth]{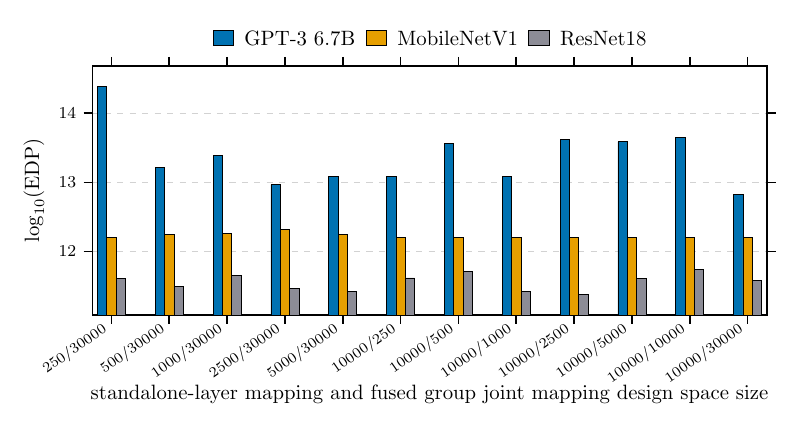}
    \caption{Effect of design space size on optimization quality. The x-axis varies the standalone and fused-group joint design space sizes, and the y-axis reports $\log_{10}(\mathrm{EDP})$.}
    \label{fig:fig8}
\end{figure}

\minisection{Effect of Design Space Size}
Continuous variables are periodically projected onto pre-enumerated discrete candidate sets whose cardinality directly shapes the optimization landscape.
\Cref{fig:fig8} sweeps candidate-set sizes for standalone layers and fused groups on Large Gemmini across three workloads.
Larger spaces generally lower EDP; the gain is most pronounced for GPT-3~6.7B---especially in the fused-group dimension---confirming that strongly coupled workloads demand a rich fusion-aware search space.
ResNet18 saturates at moderate cardinality; MobileNetV1 saturates early, its compact space well-covered even at small budgets.
Returns diminish once added candidates cease to introduce genuinely distinct solutions.

\begin{figure}[!t]
    \centering
    \includegraphics[width=1.0\columnwidth]{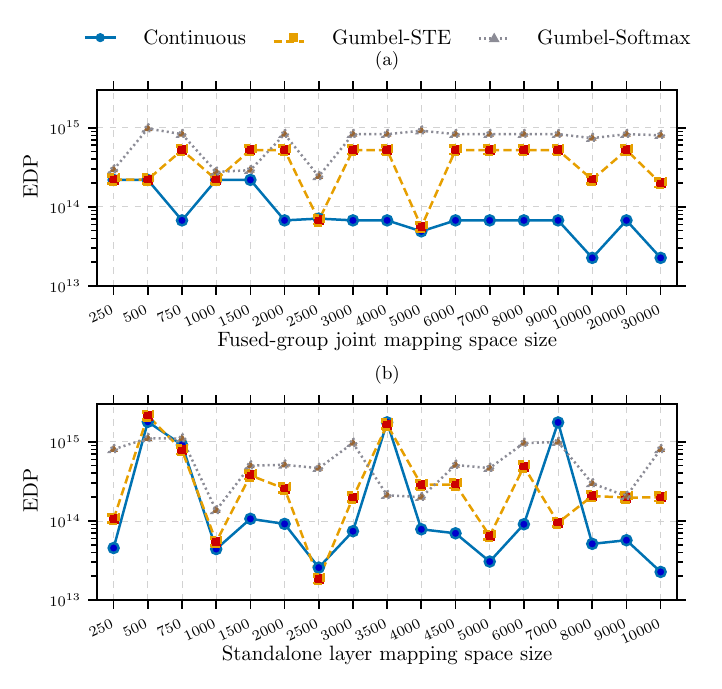}
    \caption{Discrete-optimization strategy comparison. (a)~Fused-group size swept; standalone fixed at 10{,}000. (b)~Standalone size swept; fused-group fixed at 30{,}000.}
    \label{fig:fig10}
\end{figure}

\minisection{Continuous Relaxation vs.\ Gumbel-Softmax with STE}
\label{relaxation_ablation}
We compare continuous relaxation with periodic projection, Gumbel-Softmax, and Gumbel-STE via two controlled sweeps on GPT-3~6.7B (Large Gemmini) that independently vary standalone and fused-group space sizes (\Cref{fig:fig10}).
Continuous relaxation dominates increasingly as either space grows: optimizing in a smooth landscape and projecting onto the nearest feasible point preserves local gradient structure at any cardinality.
Gumbel-Softmax degrades most severely because its weighted-average forward pass diverges from valid solutions as the categorical space expands.
Gumbel-STE mitigates this through hard forward selection but still suffers from gradient noise over large logit spaces.
The ranking is consistent: continuous relaxation $>$ Gumbel-STE $>$ Gumbel-Softmax in scalability.

%% file: doc/conclu.tex
\section{Conclusion}
\label{sec:conclusion}


We present FADiff, a differentiable framework that jointly optimizes intra-layer mapping and inter-layer fusion. 
FADiff introduces a unified, hierarchical parameterization in which continuous mapping factors and discrete fusion-template logits are co-optimized through augmented-Lagrangian penalties and Gumbel-STE selection over CP-SAT-pre-enumerated candidates. 
This formulation preserves gradient structure while guaranteeing hardware-valid solutions at every projection step. 
By reconciling continuous optimization with discrete constraints, FADiff efficiently navigates an otherwise intractable joint design space, achieving 38.8\% average EDP reduction over existing methods.